\documentstyle[prl,aps,epsfig,multicol]{revtex}
\begin{document}
\title{\bf Exact Asymptotic Results for a Model of Sequence Alignment}
\author{Satya N. Majumdar $^{1,2}$ and Sergei Nechaev $^{2,3}$}
\address{\small \it $^1$Laboratoire de Physique Theorique (UMR C5152 du CNRS), Universit\'e
Paul Sabatier, 31062 Toulouse Cedex. France \\ $^2$Laboratoire de Physique
Th\'eorique et Mod\`eles Statistiques, Universit\'e Paris-Sud. B\^at. 100. 91405
Orsay Cedex. France \\ $^3$L.D.Landau Institute for Theoretical Physics, 117334
Moscow. Russia}
\date{\today}

\maketitle

\begin{abstract}
Finding analytically the statistics of the longest common subsequence (LCS) of a
pair of random sequences drawn from $c$ alphabets is a challenging problem in
computational evolutionary biology. We present exact asymptotic results for the
distribution of the LCS in a simpler, yet nontrivial, variant of the original model
called the Bernoulli matching (BM) model which reduces to the original model in
the $c\to \infty$ limit. We show that in the BM model, for all $c$, the distribution
of the asymptotic length of the LCS, suitably scaled, is identical to the Tracy-Widom
distribution of the largest eigenvalue of a random matrix whose entries are drawn
from a Gaussian unitary ensemble. In particular, in the $c\to \infty$ limit, this
provides an exact expression for the asymptotic length distribution in the original
LCS problem.

\noindent

\medskip\noindent {PACS numbers: 87.10.+e, 87.15.Cc, 02.50.-r, 05.40.-a}
\end{abstract}

\begin{multicols}{2}

Sequence alignment is one of the most useful quantitative methods used in
evolutionary molecular biology\cite{W1,Gusfield,DEKM}. The goal of an alignment
algorithm is to search for similarities in patterns in different sequences. A
classic and much studied alignment problem is the so called `longest common
subsequence' (LCS) problem. The input to this problem is a pair of sequences
$\alpha=\{\alpha_1, \alpha_2,\dots, \alpha_i\}$ (of length $i$) and
$\beta=\{\beta_1, \beta_2,\dots, \beta_j\}$ (of length $j$). For example, $\alpha$
and $\beta$ can be two random sequences of the $4$ base pairs $A$, $C$, $G$, $T$ of
a DNA molecule, e.g., $\alpha=\{A, C, G, C, T, A, C\}$ and $\beta=\{C, T, G, A,
C\}$. A subsequence of $\alpha$ is an ordered sublist of $\alpha$ (entries of which
need not be consecutive in $\alpha$), e.g, $\{C, G, T, C\}$, but not $\{T, G, C\}$.
A common subsequence of two sequences $\alpha$ and $\beta$ is a subsequence of both
of them. For example, the subsequence $\{C, G, A, C\}$ is a common subsequence of
both $\alpha$ and $\beta$. There can be many possible common subsequences of a pair
of sequences. The aim of the LCS problem is to find the longest of such common
subsequences. This problem and its variants have been widely studied in
biology\cite{NW,SW,WGA,AGMML}, computer science\cite{SK,AG,WF,Gusfield}, probability
theory\cite{CS,Deken,Steele,DP,Alex,KLM} and more recently in statistical
physics\cite{ZM,Hwa,Monvel}. A particularly important application of the LCS problem
is to quantify the closeness between two DNA sequences. In evolutionary biology, the
genes responsible for building specific proteins evolve with time and by finding the
LCS of the same gene in different species, one can learn what has been conserved in
time. Also, when a new DNA molecule is sequenced {\it in vitro}, it is important to
know whether it is really new or it already exists. This is achieved quantitatively
by measuring the LCS of the new molecule with another existing already in the
database.

For a pair of fixed sequences of length $i$ and $j$ respectively, the length
$L_{i,j}$ of their LCS is just a number. However, in the stochastic version of the
LCS problem one compares two random sequences drawn from $c$ alphabets and hence the
length $L_{i,j}$ is a random variable. A major challenge over the last three decades
has been to determine the statistics of $L_{i,j}$\cite{CS,Deken,Steele,DP,Alex}. For
equally long sequences ($i=j=n$), it has been proved that $\langle L_{n,n}\rangle
\approx \gamma_c n$ for $n\gg 1$, where the averaging is performed over all
realizations of the random sequences. The constant $\gamma_c$ is known as the
Chv\'atal-Sankoff constant which, to date, remains undetermined though there exists
several bounds\cite{Deken,DP,Alex}, a conjecture due to Steele\cite{Steele} that
$\gamma_c=2/(1+\sqrt{c})$ and a recent proof\cite{KLM} that $\gamma_c\to 2/\sqrt{c}$
as $c\to \infty$. Unfortunately, no exact results are available for the finite size
corrections to the leading behavior of the average $\langle L_{n,n}\rangle$, for the
variance, and also for the full probability distribution of $L_{n,n}$. Thus, despite
tremendous analytical and numerical efforts, exact solution of the random LCS
problem has, so far, remained elusive. Therefore it is important to find other
variants of this LCS problem that may be analytically tractable.

Computationally, the easiest way to determine the length $L_{i,j}$ of the LCS of two
arbitrary sequences of lengths $i$ and $j$ (in polynomial time $\sim O(ij)$) is via
using the recursive algorithm\cite{Gusfield,Monvel}
\begin{equation}
L_{ij} = \max\left[L_{i-1,j}, L_{i,j-1}, L_{i-1,j-1} + \eta_{i,j}\right],
\label{recur1}
\end{equation}
subject to the initial conditions $L_{i,0}=L_{0,j}=L_{0,0}=0$. The variable
$\eta_{i,j}$ is either 1 when the characters at the positions $i$ (in the sequence
$\alpha$) and $j$ (in the sequence $\beta$) match each other, or 0 if they do not.
Note that the variables $\eta_{i,j}$'s are not independent of each other. To see
this consider the simple example -- matching of two strings $\alpha={\rm AB}$ and
$\beta={\rm AA}$. One has by definition: $\eta_{1,1}=\eta_{1,2}=1$ and
$\eta_{2,1}=0$. The knowledge of these three variables is sufficient to predict that
the last two letters will not match, i.e., $\eta_{2,2}=0$. Thus, $\eta_{2,2}$ can
not take its value independently of $\eta_{1,1},\,\eta_{1,2},\,\eta_{2,1}$. These
residual correlations between the $\eta_{i,j}$ variables make the LCS problem rather
complicated. Note however that for two random sequences drawn from $c$ alphabets,
these correlations between the $\eta_{i,j}$ variables vanish in the $c\to \infty$
limit.

A simpler but natural variant of this LCS problem is the Bernoulli matching (BM)
model where one ignores the correlations between $\eta_{i,j}$'s for all
$c$\cite{Monvel}. The BM model reduces to the original LCS problem only in the $c\to
\infty$ limit. The length $L_{i,j}^{BM}$ of the BM model satisfies the same
recursion relation in Eq. (\ref{recur1}) except that $\eta_{i,j}$'s are now
independent and each drawn from the bimodal distribution: $p(\eta)=
(1/c)\delta_{\eta,1}+ (1-1/c)\delta_{\eta,0}$. The BM model, though simpler than the
original LCS problem, is still nontrivial due to the nonlinear recursion relation in
Eq. (\ref{recur1}). Using the cavity method of spin glass physics\cite{MPV}, the
asymptotic behavior of the average length in the BM model was determined
analytically\cite{Monvel},
\begin{equation}
\langle L_{n,n}^{BM}\rangle  \approx \gamma_c^{BM} n
\label{bm1}
\end{equation}
where $\gamma_c^{BM}= 2/(1+\sqrt{c})$, same as the conjectured value of the
Chv\'atal-Sankoff constant $\gamma_c$ for the original LCS model. However, other
properties such as the variance or the distribution of $L_{n,n}^{BM}$ remained
untractable even in the BM model.

The purpose of this Letter is to present an exact asymptotic formula for the
distribution of the length $L_{n,n}^{BM}$ in the BM model for all $c$. Our main
result is that for large $n$,
\begin{equation}
L_{n,n}^{BM}\to \gamma_c^{BM} n + f(c)\, n^{1/3}\, \chi \label{asymp1}
\end{equation}
where $\chi$ is a random variable with a $n$-independent distribution, ${\rm Prob}
(\chi\le x)= F_{\rm TW}(x)$ which is the well studied Tracy-Widom distribution for
the largest eigenvalue of a random matrix with entries drawn from a Gaussian unitary
ensemble\cite{TW}. For a detailed form of the function $F_{\rm TW}(x)$, see
\cite{TW}. We show that for all $c$,
\begin{equation}
f(c)=\frac{c^{1/6}(\sqrt{c}-1)^{1/3}}{\sqrt{c}+1}.
\label{fc1}
\end{equation}
This allows us to calculate the average including the subleading finite size
correction term and the variance of $L_{n,n}^{BM}$ for large $n$,
\begin{eqnarray}
\langle L_{n,n}^{BM}\rangle &\approx & \gamma_c^{BM} n + \left<\chi\right> f(c)
n^{1/3} \nonumber \\
{\rm Var}\, L_{n,n}^{BM} &\approx &
\left(\langle\chi^2\rangle-{\langle\chi\rangle}^2\right)\, f^2(c)\, n^{2/3},
\label{eq:expvar}
\end{eqnarray}
where one can use the known exact values\cite{TW}, $\langle \chi\rangle=
-1.7711\dots$ and $\langle \chi^2\rangle- {\langle \chi\rangle}^2= 0.8132\dots$. In
particular, we note that in the limit $c\to \infty$, Eqs.
(\ref{asymp1})-(\ref{eq:expvar}) provide
exact asymptotic results for the original LCS model as well.

In the BM model, the length $L_{i,j}^{BM}$ can be interpreted as the height of a
surface over the $2$-d $(i,j)$ plane constructed via the recursion relation in Eq.
(\ref{recur1}). A typical surface, shown in Fig. (1a), has terrace-like structures.
\begin{figure}[ht]
\centerline{\epsfig{file=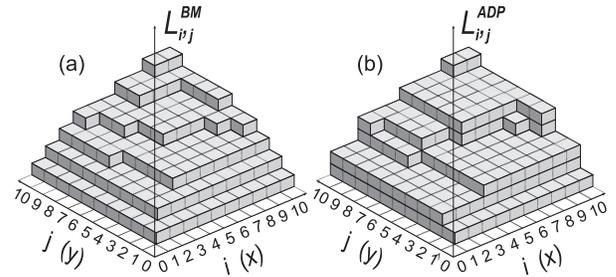,width=8cm}}
\caption{Examples of (a) BM surface
$L_{i,j}^{BM}\equiv {\tilde h}(x,y)$ and (b) ADP surface $L_{i,j}^{ADP}\equiv
h(x,y)$.} \label{fig:1}
\end{figure}

It is useful to consider the projection of the level lines separating the adjacent
terraces whose heights differ by $1$ (see Fig.2) onto the $2$-d $(i,j)$ plane. Note
that, by the rule Eq. (\ref{recur1}), these level lines never overlap each other,
i.e., no two paths have any common edge. The statistical weight of such a projected
$2$-d configuration is the product of weights associated with the vertices of the
$2$-d plane. There are five types of possible vertices with nonzero weights as shown
in Fig.2, where $p=1/c$ and $q=1-p$. Since the level lines never cross each other,
the weight of the first vertex in Fig. (2) is 0.
\begin{figure}[ht]
\centerline{\epsfig{file=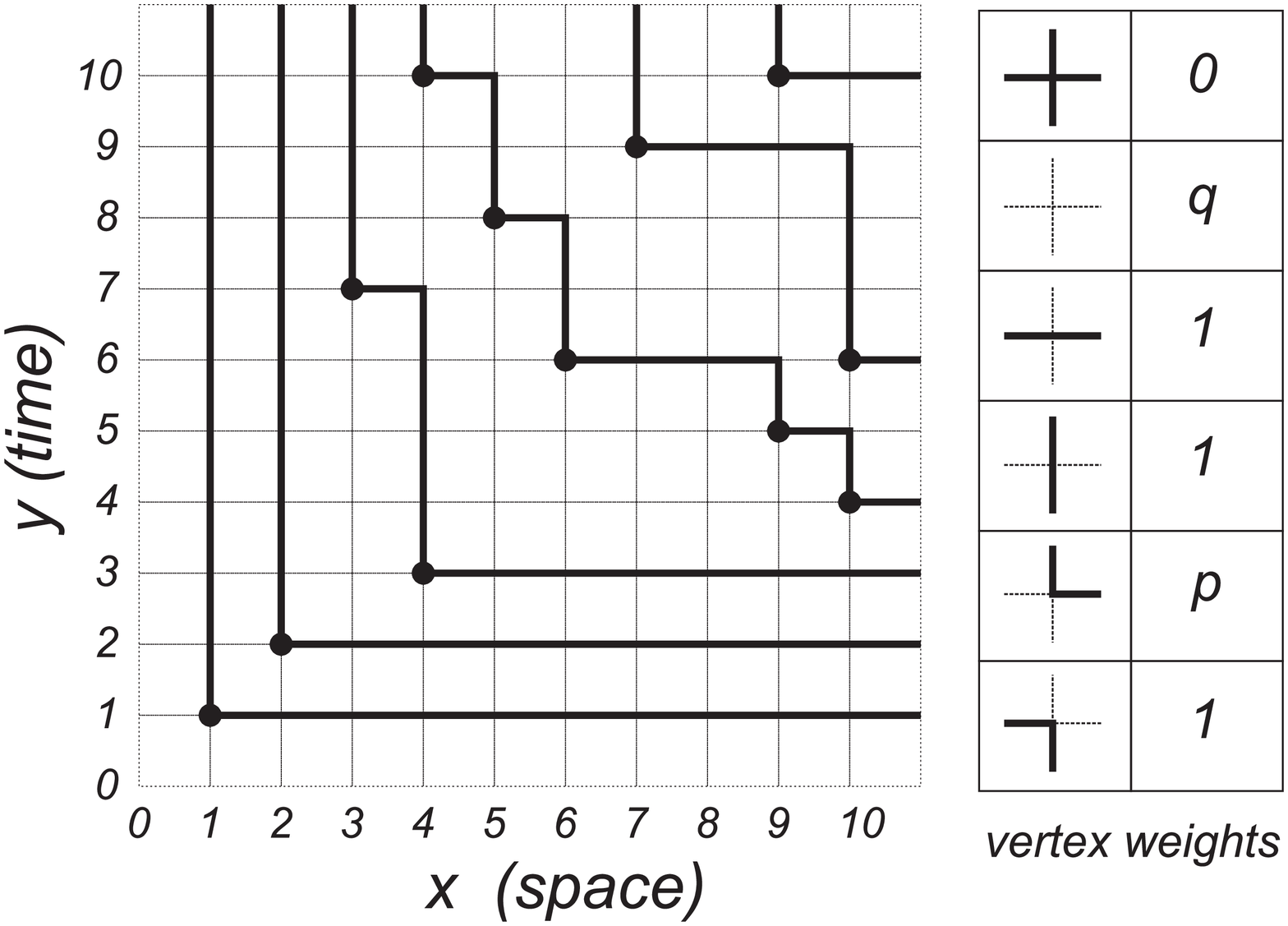,width=6.5cm}}
\caption{Projected $2$-d level lines separating adjacent terraces of unit height
difference in the BM surface in Fig.(1a). The adjacent table shows the weights of
all vertices on the $2$-d plane.} \label{fig:2}
\end{figure}

Consider first the limit $c\to \infty$ (i.e., $p\to 0$). The weights of all allowed
vertices are $1$, except the ones shown by black dots in Fig.(2), whose associated
weights are $p\to 0$. The number $N$ of these black dots inside a rectangle of area
$A=ij$ can be easily estimated. For large $A$ and $p\to 0$, this number is Poisson
distributed with the mean ${\overline N}= pA$. The Bethe ansatz analysis shows that
BM corresponds to the sector of the 5-vertex model\cite{Wu} where the density
$\alpha$ of empty edges in a row of vertical edges is close to the boundary
$\alpha\approx 1^{-}$. The careful examination of the free energy near this boundary
allows one to conclude that the leading contribution in $p$ (for $p\to 0$) to
${\overline N}$ comes exactly from the line of phase transitions in a 5-vertex
model. The subleading corrections to ${\overline N}$ are of order $\sim p^{3/2}$ and
are ensured by small deviations from the critical line being beyond the Poisson
approximation\cite{MN}.

The height $L_{i,j}^{BM}$ is just the number of level lines $\cal N$ inside this
rectangle of area $A=ij$. The problem of estimating $\cal N$ has recently appeared
in a number of interface models such as a polynuclear growth model\cite{PS} and a
ballistic deposition model\cite{BD}. By using a mapping to the longest increasing
subsequence (LIS) of the equally likely permutations of a set of integers and then,
by applying a celebrated result due to Baik, Deift and Johansson (BDJ)\cite{BDJ}, it
was shown\cite{PS,BD} that the number of level lines ${\cal N}$ inside the rectangle
(for large $A$), appropriately scaled, has a limiting behavior, ${\cal N}\to
2\sqrt{\overline N} + {\overline N}^{1/6}\, \chi$, where $\chi$ is a random variable
with Tracy-Widom distribution. Using ${\overline N}=pA=ij/c$, one then obtains in
the limit $p\to 0$,
\begin{equation}
L_{i,j}^{BM}= {\cal N} \to \frac{2}{\sqrt c}\sqrt{ij} +
{\left( \frac{ij}{c}\right)}^{1/6}\, \chi.
\label{p01}
\end{equation}
In particular, for large equal length sequences $i=j=n$, we get for $c\to \infty$
\begin{equation}
L_{n,n}^{BM}\to \frac{2}{\sqrt{c}}\, n + c^{-1/6} \, n^{1/3}\, \chi .
\label{p02}
\end{equation}
Note that since the BM and the original LCS model are equivalent in the limit $c\to
\infty$, the exact results in Eqs. (\ref{p01})-(\ref{p02}) also hold for the LCS
model. Note that only the leading behavior of the average $\langle L_{n,n}\rangle$
was known before\cite{KLM} in the $c\to \infty$ limit of the original LCS model.

For finite $c$, while the above mapping to the LIS problem still works, the
corresponding permutations of the LIS problem are not generated with equal
probability and hence one can no longer use the BDJ results. To make progress for
finite $c$, we map the BM model exactly to a $3$-d anisotropic directed percolation
(ADP) model first considered by Rajesh and Dhar\cite{RD}. This ADP model can further
be mapped to a $(1+1)$-d directed polymer problem studied by
Johansson\cite{Johansson}. For this specific directed polymer problem, Johansson
derived exact asymptotic result for the distribution of the polymer energy.
Translating these results back to the BM model, we derive our main results in Eqs.
(\ref{asymp1})-(\ref{eq:expvar}). Note that the recursion relation in Eq.
(\ref{recur1}) can also be viewed as a $(1+1)$-d directed polymer
problem\cite{Hwa,Monvel} and some asymptotic results (such as the $O(n^{2/3})$
behavior of the variance of $L_{n,n}$ for large $n$) can be obtained using the
arguments of universality\cite{Hwa}. However, this does not provide precise results
for the full distribution which are obtained here.

Let us consider a directed bond percolation on a simple cubic lattice. The bonds are
occupied with probabilities $p_x$, $p_y$, and $p_z$ along the $x$, $y$ and $z$ axes
and are all directed towards increasing coordinates. Imagine a source of fluid at
the origin which spreads along the occupied directed bonds. The sites that get wet by the
fluid form a $3$-d cluster. In the ADP problem, the bond occupation probabilities are
anisotropic, $p_x=p_y=1$ (all bonds aligned along the $x$ and $y$ axes are occupied)
and $p_z=p$. Hence, if the point $(x,y,z)$ gets wet by the fluid then all the points
$(x',y', z)$ on the same plane with $x'\ge x$ and $y'\ge y$ also get wet. Such a wet
cluster is compact and can be characterized by its bounding surface height $h(x,y)$
as shown in Fig.(1b). It is not difficult to see that the height $h(x,y)$ satisfies
the following recursion relation\cite{RD},
\begin{equation}
h(x,y) = \max \left[ h(x-1,y), h(x, y-1)\right] + \xi_{i,j},
\label{recur2}
\end{equation}
where $\xi_{i,j}$'s are i.i.d. random variables taking nonnegative integer values
with ${\rm Prob}(\xi_{i,j}=k)= (1-p)\, p^k$ for $k=0,1,2,\dots$. One can also
interpret the height $h(x,y)$ in Eq. (\ref{recur2}) as the energy of a directed
polymer in the $(x-y)$ plane. Precisely this particular version of the polymer
problem was studied by Johansson\cite{Johansson} who obtained the asymptotic
distribution of the height for large $x$ and $y$,
\begin{eqnarray}
h(x,y) &\to& \frac{2\sqrt{pxy}+p(x+y)}{q}+ \nonumber \\
       &+&   \frac{(pxy)^{1/6}}{q}\,\left[(1+p)+\sqrt{\frac{p}{xy}}\,(x+y)\right]^{2/3}
       \, \chi,
\label{j1}
\end{eqnarray}
where $q=1-p$, $\chi$ is a random variable with a Tracy-Widom distribution.

While the terrace-like structures of the ADP surface look similar to the BM surfaces
(compare Figs.(1a) and (1b)), there is an important difference between the two. In
the ADP model, the level lines separating two adjacent terraces can overlap with
each other\cite{RD}, which does not happen in the BM model. However, by making the
following change of coordinates in the ADP model\cite{RD}
\begin{equation}
\zeta= x+ h(x,y); \,\,\, \eta=y+ h(x,y)
\label{ct1}
\end{equation}
one gets a configuration of the surface where the level lines no longer overlap.
Moreover, it is not difficult to show that the projected $2$-d configuration of
level lines of this shifted ADP surface has exactly the same statistical weight as
the projected $2$-d configuration of the BM surface. Denoting the BM height by
${\tilde h}(x,y)= L_{x,y}^{BM}$, one then has the identity, ${\tilde h}(\zeta,
\eta)= h(x,y)$, which holds for each configuration. Using Eq. (\ref{ct1}), one can
rewrite this identity as
\begin{equation}
{\tilde h}(\zeta, \eta)= h\left( \zeta- {\tilde h}(\zeta, \eta),
\eta- {\tilde h}(\zeta, \eta)\right).
\label{conv1}
\end{equation}
Thus, for any given height function $h(x,y)$ of the ADP model, one can, in
principle, obtain the corresponding height function ${\tilde h}(x,y)$ for all
$(x,y)$ of the BM model by solving the nonlinear equation (\ref{conv1}). This is
however very difficult in practice. Fortunately, one can make progress for large
$(x,y)$ where one can replace the integer valued discrete heights by continuous
functions $h(x,y)$ and ${\tilde h}(x,y)$. Using the notation $\partial_x\equiv
\partial/{\partial x}$ it is easy to derive from Eq. (\ref{ct1}) the following pair
of identities,
\begin{equation}
\partial_x h = \frac{\partial_{\zeta} {\tilde h}}{1-\partial_{\zeta}
{\tilde h}-\partial_{\eta} {\tilde h}};
\,\,\,
\partial_y h = \frac{\partial_{\eta} {\tilde h}}{1-\partial_{\zeta}
{\tilde h}-\partial_{\eta} {\tilde h}}.
\label{der1}
\end{equation}
In a similar way, one can show that
\begin{equation}
\partial_{\zeta} {\tilde h} = \frac{\partial_x h}{1+\partial_x h+\partial_y h};\,\,\,
\partial_{\eta} {\tilde h} = \frac{\partial_y h}{1+\partial_x h+\partial_y h}.
\label{der2}
\end{equation}
We then observe that Eqs. (\ref{der1}) and (\ref{der2}) are invariant under the
simultaneous transformations
\begin{equation}
\zeta\to -x ; \,\, \eta\to -y; \,\, \tilde h \to h \, .
\label{invar1}
\end{equation}
Since the height is built up by integrating the derivatives, this leads to a simple
result for large $\zeta$ and $\eta$,
\begin{equation}
{\tilde h}(\zeta, \eta) = h(-\zeta, -\eta).
\label{res1}
\end{equation}
Thus, if we know exactly the functional form of the ADP surface $h(x,y)$, then the
functional form of the BM surface ${\tilde h}(x,y)$ for large $x$ and $y$ is simply
obtained by ${\tilde h}(x,y)=h(-x,-y)$. Changing $x\to -x$ and $y\to -y$ in
Johansson's expression for the ADP surface in Eq. (\ref{j1}) we thus arrive at our
main asymptotic result for the BM model
\begin{eqnarray}
L_{x,y}^{BM}&=& {\tilde h}(x,y) \to \frac{2\sqrt{pxy}-p(x+y)}{q}+ \nonumber \\
&+&\frac{(pxy)^{1/6}}{q}\,\left[(1+p)-\sqrt{\frac{p}{xy}}\,(x+y)\right]^{2/3} \,
\chi, \label{res2}
\end{eqnarray}
where $p=1/c$ and $q=1-1/c$. For equal length sequences $x=y=n$, Eq. (\ref{res2})
then reduces to Eq. (\ref{asymp1}).

To check the consistency of our asymptotic results, we further computed the
difference between the left- and the right-hand sides of Eq. (\ref{conv1}),
\begin{equation}
\Delta h (\zeta, \eta)= {\tilde h}(\zeta, \eta)- h\left( \zeta- {\tilde h}(\zeta,
\eta), \eta- {\tilde h}(\zeta, \eta)\right), \label{conv2}
\end{equation}
with the functions $h(x,y)$ and ${\tilde h}(x,y)$ given respectively by Eqs.
(\ref{j1}) and (\ref{res2}). For large $\zeta=\eta$ one gets
\begin{equation}
\Delta h(\zeta,\zeta) \to \left[{p^{1/3}\chi^2}/{3 (1-\sqrt{p})^{4/3}}\right]\,
{\zeta}^{-1/3} . \label{cons1}
\end{equation}
Thus the discrepancy falls off as a power law for large $\zeta$, indicating that
indeed our solution is asymptotically exact. We have also performed numerical
simulations of the BM model using the recursion relation in Eq. (\ref{recur1}) for
$c=2,\,4,\,9,\,16,\,100$. Our preliminary results\cite{MN} for relatively small
system sizes (up to $n=5000$) are consistent with our exact results in Eqs.
(\ref{asymp1})-(\ref{eq:expvar}).

The Tracy-Widom distribution of the random matrix theory has appeared recently in a
number of problems\cite{TW,AD,Johansson,PS,BD}. In this Letter, we have shown that
it also describes the asymptotic distribution of the length of the longest common
subsequence in a sequence matching problem. While a possible link
between the two problems was speculated before\cite{AD}, a precise
connection, so far, was missing and is provided here.

\end{multicols}

\end{document}